\documentclass[a4paper]{article}

\usepackage{INTERSPEECH2021}
\usepackage{cite}
\usepackage{amsmath}

\title{Learning Metrics from Mean Teacher: A Supervised Learning Method for Improving the Generalization of Speaker Verification System}
\name{Ju-ho Kim$^1$, Hye-jin Shim$^1$, Jee-weon Jung$^{1,2}$, and Ha-Jin Yu$^1{\dag}$\thanks{$^\dag$ Corresponding author}}
\address{$^1$School of Computer Science, University of Seoul, $^2$Naver Corporation}
\email{wngh1187@naver.com, shimhz6.6@gmail.com, jeeweon.jung@navercorp.com, hjyu@uos.ac.kr}

\begin{document}

\maketitle

\begin{abstract}
Most speaker verification tasks are studied as an open-set evaluation scenario considering the real-world condition. 
Thus, the generalization power to unseen speakers is of paramount important to the performance of the speaker verification system. 
We propose to apply \textit {Mean Teacher}, a temporal averaging model, to extract speaker embeddings with small intra-class variance and large inter-class variance. 
The mean teacher network refers to the temporal averaging of deep neural network parameters; it can produces more accurate and stable representations than using weights after the training finished. 
By learning the reliable intermediate representation of the mean teacher network, we expect that the proposed method can explore more discriminatory embedding spaces and improve the generalization performance of the speaker verification system. 
Experimental results on the VoxCeleb1 test set demonstrate that the proposed method relatively improves performance by 11.61\%, compared to a baseline system. 
\end{abstract}

\section{Introduction}
A speaker verification (SV) task authenticates whether a speaker of an unknown input utterance matches the target speaker, and is widely utilized in applications such as a voice assistant system \cite{hansen2015speaker, bai2020speaker}. 
The SV task is mainly studied as an open-set scenario that tests using the unseen speaker's utterances during training phase, which requires strong generalization performance of the model \cite{bai2020speaker, bendale2016towards}. 
Several recent studies have trained deep neural networks (DNNs) using metric learning-based objective functions rather than classification-based objective functions by considering the characteristics of SV task; these systems have demonstrated outstanding performance in terms of open-set SV evaluation \cite{kye2020meta, chung2020defence}. 

In \cite{polyak1992acceleration}, the authors show that by solely averaging the DNN's weight parameters after each step in the training phase can lead to more stable and accurate results than using the final weights directly, and refer to this technique as ``temporal averaging''. 
By leveraging this knowledge, in the field of semi-supervised learning, Tarvainen \textit{et al.} \cite{tarvainen2017mean} proposed a novel framework that utilizes a temporal averaging model, called \textit{Mean Teacher} (MT), to train the DNN using unlabeled data. 
The MT framework comprises a teacher–student \cite{hinton2015distilling} setting, wherein the teacher network, i.e, MT, is updated with an exponential moving average (EMA) of a set of student network parameters at each training step. 
Thus, the student network can learn unlabeled data by reducing the Euclidean distance from the MT network's relatively accurate predictions. 

The MT network can be regarded as a temporal ensemble model in terms of aggregating the information about the student network at each training step. 
This MT network provides high-quality features for the student network to learn stably. 
From this perspective, we hypothesize that the MT network as a kind of ensemble teacher \cite{freitag2017ensemble} can be a sufficient reference model to support student network training even in the supervised learning domain. 
Consequently, in this study, we propose a method that can improve the generalization performance of open-set SV task by adapting the MT framework in a supervised learning condition. 

To widen the representation space and improve the performance of the SV system, we modify the following two factors in the existing MT framework. 
First, the student network directly learns the speaker embeddings output from the MT network instead of the predictions. 
Second, the consistency loss function between the student and the MT network is changed to a cosine similarity-based metric learning that utilizes the negative pairs together, rather than mean square error (MSE), which considers only positive pairs. 
All experiments performed herein use the entire VoxCeleb2 \cite{chung2018voxceleb2} dataset as a training dataset and the VoxCeleb1 \cite{nagrani2017voxceleb} test set as an evaluation dataset. 
As a result of the experiment, the proposed method demonstrates a relative error reduction (RER) of 11.61\% compared to the baseline system. 

\section{Related work}
\label{sec:related_works}

\textbf{DNN-based open-set speaker verification:} 
Many of the early works on DNN-based SV have usually adopted classification-based objective functions such as softmax \cite{variani2014deep, snyder2018x}. 
However, open-set SV requires a discriminative embedding space for unseen speakers rather than accurate classification on training datasets. 
Therefore, several studies have utilized a variant of the softmax loss function applying an additional angular margin to reduce the variance within the class \cite{wang2018cosface, deng2019arcface}. 
In contrast, open-set SV can essentially be treated as a metric learning problem \cite{chung2020defence, bai2020speaker}. 
Thus, many recent studies have improved the generalization performance by optimizing metric learning-based cost functions rather than classification-based cost functions \cite{snell2017prototypical, wan2018generalized}. 
In addition to the aforementioned loss functions, preceding studies have investigated diverse methods such as data augmentation \cite{zhang2017mixup, park2019specaugment}, network architectures \cite{snyder2017deep, he2016deep}, and system frameworks \cite{sang2020open, tao2020improving}. 

\textbf{Temporal averaging and mean teacher:} 
Loss values oscillating or bouncing without convergence during DNN training is a common training failure indicator \cite{vogl1988accelerating}. 
To alleviate this issue, Polyak \textit{et al.} \cite{polyak1992acceleration} propose a temporal averaging method that combines the weights collected until the end of the training. 
In the loss landscape, temporal averaging brings it closer to the bottom of the valley by averaging the weights of the points that oscillate back and forth \cite{goodfellow2016deep}. 
Therefore, this method can be used to improve the convergence of an optimization algorithm or to further enhance the generalization performance of the model during evaluation \cite{kingma2014adam, szegedy2016rethinking}. 

The MT framework using a temporal averaging model has been proposed for semi-supervised image classification tasks \cite{tarvainen2017mean}. 
The architectures of the student and MT networks are identical, and the parameters of the student network $\theta$ are trained via back-propagation; the weights of the MT network $\xi$ are updated through the EMA of student weights, formulated as: 
\begin{equation} 
\label{eq1}
    \xi_t = \alpha\xi_{t-1} + (1-\alpha)\theta_t
\end{equation} 
where $t$ indicates the training step and $\alpha$ is a smoothing coefficient hyper-parameter. 

The same mini-batch is fed to the student and MT networks, but different noises such as dropout or augmentation are added to each input. 
The loss function for the student network is a linear combination of classification and consistency costs. 
The classification cost is only applied to labeled data and adopts a categorical cross-entropy (CCE) function. 
For all data, including unlabeled data, the consistency loss function (e.g., MSE) is constrained to reduce the Euclidean distance between the prediction of the student and MT networks. 
The key point of the MT framework is to form an MT network that progressively aggregates information from the student network in an EMA fashion. 
Consequently, the MT network is expected to produce more reliable probability predictions that serve high-quality representation to guide the student network’s training. 

\section{Baseline}
\label{sec:baseline}
Recently, based on the assumption that hand-crafted features may not be optimal, a data-driven systems that are fed by less processed features such as spectrograms or raw waveforms are being explored in audio domains \cite{nagrani2017voxceleb, badshah2017speech, fu2017raw, ravanelli2018speaker}. 
In the SV field, a model that is directly fed by raw waveforms is hypothesized that the first one-dimensional convolution layer can appropriately aggregate various frequency bands of utterances and potentially extract discriminative features \cite{ravanelli2018speaker, jung2019rawnet}. 
Therefore, we use raw waveforms as input in all experiments and utilize RawNet2 \cite{jung2020improved, jung2020alpha}, a representative neural speaker embedding extractor based on raw waveform input, as the baseline system. 

Table \ref{tab:DNN_arch} describes the structure of the baseline with several modifications made to improve the performance of the original RawNet2. 
First, we increase the number of residual blocks \cite{he2016deep} in the model from six to eight. 
Second, instead of the gated-recurrent unit, we employ attentive statistics pooling \cite{okabe2018attentive} to aggregate frame-level features into an utterance level feature. 
Third, we reduce the embedding dimension from 1,024 to 512. 
More specific details related to the system architecture are described in the literature \cite{jung2020alpha} and the author's GitHub\footnote{https://github.com/Jungjee/RawNet}.

\begin{table}[t!]
 \caption{DNN architecture of the baseline system. }
  \centering
  \label{tab:DNN_arch}
  \begin{tabular}{l c c c c}
  \toprule 
  \textbf{Layer} & \textbf{Kernel size} & \textbf{Stride} & \textbf{\# Block} & \textbf{Output shape}\\
  \toprule
  1D-Conv & 3 &  3 & 1 & L/3 $\times$ 128\\
  \midrule
  Res1 & 3 $\times$ 3 & 1 $\times$ 1  & 2 & L/$3^3$ $\times$ 128\\
  \midrule
  Res2 & 3 $\times$ 3 & 1 $\times$ 1 & 3 & L/$3^6$ $\times$ 256\\
  \midrule
  Res3 & 3 $\times$ 3 & 1 $\times$ 1 & 3 & L/$3^9$ $\times$ 512\\
  \midrule
  ASP & - & - & 1 &1024\\
  \midrule
  Linear & 512 &- & 1 &512\\
  \bottomrule
  \end{tabular}
\end{table}

\begin{table}[t]
    \caption{The performance of the original system and the modified version of RawNet2.}
    \label{tab:baseline}
    \centering
    \begin{tabular}{l|c|c}
        \Xhline{1pt} 
        System & Batch size & EER \\
        \Xhline{1pt} 
        \#1-RawNet2 \cite{jung2020alpha} & 120 & 2.31\\
        \textbf{\#2-Baseline} &200&\textbf{2.24}\\
        \Xhline{1pt} 
    \end{tabular}
    \vskip -5pt
\end{table}

Table \ref{tab:baseline} shows the performance of the original RawNet2 and the baseline. 
Through this experiment, our baseline model reported better performance than the original RawNet2 with an equal error rate (EER) of 2.24\%.

\section{Proposed method}
\label{sec:proposed}

\subsection{Architecture}
\begin{figure}[t]
  \centering
  \includegraphics[width=\linewidth]{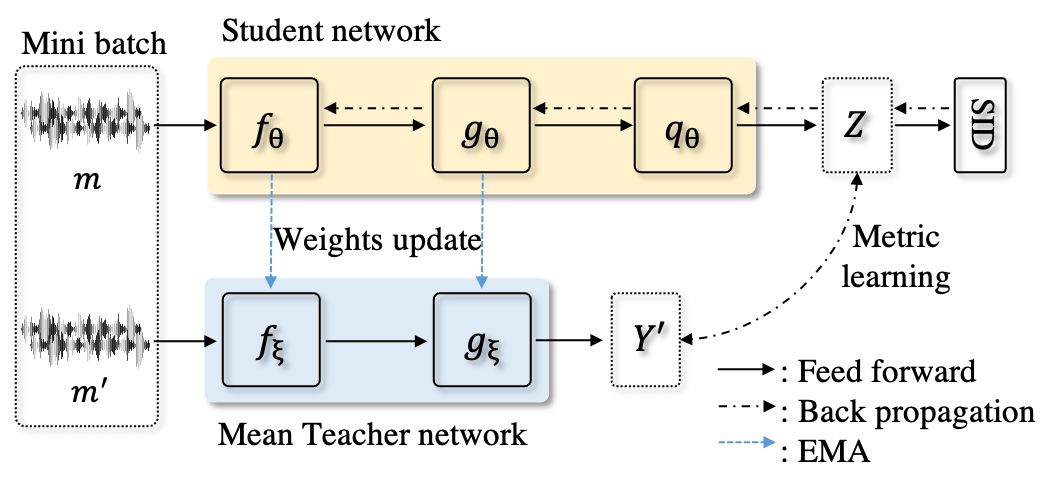}
  \vskip -3pt
  \caption{Overall structure of the proposed method. 
  The student and mean teacher (MT) networks comprise an encoder $f$ and projector $g$, and the predictor $q$ is added only to the student. 
  The two networks are fed by separate mini-batch $m$ and $m'$ and extract the embeddings $Z$ and $Y'$, respectively. 
  The student network is trained by cosine similarity-based metric learning using $Y'$, along with speaker identification. 
  The parameter set of the MT network $\xi$ is updated with the exponential moving average (EMA) of the student network parameter set $\theta$. 
  }
  \label{fig:proposed_framework}
  \vskip -5pt
\end{figure}

We aim to achieve a better generalization performance toward open-set SV scenarios. 
Assuming that a network with temporal averaging weights that yields more accurate and stable results can be a sufficient reference model, we propose to adapt the MT framework to SV. 
Figure \ref{fig:proposed_framework} shows the overall structure of the proposed method. 
Two networks, student and MT, comprise an encoder $f$ and projector $g$, wherein the predictor $q$ is added only to the student network. 
This asymmetric structure is designed with inspiration from \cite{grill2020bootstrap}: the predictors of the student network can explore a different embedding space than MT network. 
$\theta$ and $\xi$ indicate the set of parameters for the student and MT networks, respectively. 
The encoder extracts the intermediate representation from the input utterances, and we designate it as RawNet2, which is used as a baseline in this study. 
The projector and predictor project the embeddings into different representation spaces. 
These are identical structure to each other, comprising a batch normalization \cite{ioffe2015batch} and a rectified linear unit \cite{nair2010rectified} between two fully-connected layers with 512 nodes. 

\subsection{Feed-forward}
One of the ways to learn a manifold of unlabeled data in a semi-supervised or self-supervised learning domain is to reduce the distance between predictions after applying different augmentation or noise \cite{tarvainen2017mean, grill2020bootstrap}. 
However, in the field of supervised learning in which all labels exist, it is more effective to fully utilize the labels rather than the indirect learning method descried above. 
In addition, the ultimate goal of an SV task is not to classify but to compare two input utterances in the representation space. 
Thus, herein, we intend to design a model to directly decrease the within-class variance by explicitly reducing the distance between embeddings output from different data of the same class, i.e., different utterances of the same speaker. 
Let $M$ be a mini-batch, $M \in \mathbb{R}^{S\times U\times T}$ where $S$, $U$, and $T$ refer to the number of speakers, the number of utterances of a speaker, and the length of a sample, respectively. 
$M$ is divided into $m$ and $m'$; each of $m$ and $m'$ has half of the utterances of the speakers, where $m, m' \in \mathbb{R}^{S\times {U\over 2}\times T}$. 
Therefore, $m$ and $m'$ comprise the pairs of different utterance sets for each speaker, which are fed to the student and MT networks and the two networks extract the embeddings $Z$ and $Y'$, respectively. 

\begin{equation}
    Z = q_\theta(g_\theta(f_\theta(m))),\quad Y' =g_\xi(f_\xi(m'))
\end{equation}
where $Z, Y' \in \mathbb{R}^{S\times {U\over 2}\times D}$ and $D$ is the embedding dimension (designated as 512 in this study). 
After training is completed, SV performance is evaluated using the embeddings $Z$ extracted from the student network. 

\subsection{Updating the mean teacher network}
\label{MT}
In this study, we expect that the temporal averaging weight model, i.e., MT, can be a sufficient reference model to support student network training in terms of being a kind of ensemble teacher. 
The MT network parameter set $\xi$ is updated using the EMA of the student network parameter set $\theta$ as in eq. (\ref{eq1}).  
We set the smoothing coefficient hyper-parameter to 0.99. 
This value implies that the MT network linearly reduces the weight of the past student network, which is far from optimal and aggregates the latest information. 
Consequently, we hypothesize that the MT network can provide stable and accurate speaker embeddings to train the student network. 

\subsection{Student network training}
\label{Student}
Unlike the MT network, the weights of the student network are updated by the back-propagation algorithm. 
There are two loss functions for the student network: consistency and classification costs. 
First, to discriminatively project the utterances of unseen speakers to the representation space, it is necessary to learn not only to reduce intra-class variance, but also to increase inter-class variance. 
Thus, we designate the consistency cost as a cosine similarity-based metric learning loss function using embeddings from the two networks, rather than the MSE between predictions, which considers only positive pairs as in a previous study \cite{tarvainen2017mean}. 
It is expected that the embedding space can be widened by using a negative pair as well as a positive pair between the two network embeddings. 
In addition, by utilizing the cosine similarity metric, which is a measure of similarity between embeddings in SV evaluation, it is possible to directly learn the embeddings suitable for SV. 

We use the centroid-based metric learning loss function, such as generalized end-to-end (GE2E) \cite{wan2018generalized} and angular prototypical (AP) \cite{chung2020defence}. 
In the case of GE2E, we modified it to fit our proposed method. 
The original GE2E loss operates using cosine similarity between each query embedding and the centroids in the mini-batch. 
Herein, the query refers to each utterance of all speakers, and the centroid $c$ is derived by averaging embeddings that belong to the same speaker. 
Note that if speakers of a query and the centroid are equal, the centroid is calculated excluding that query. 
In the proposed method, only the student network is back-propagated; hence, only the embedding of the student network $Z$ is used as a query during training and is referred to as half GE2E (GE2E-H). 

\begin{equation}
    c_j = \frac{1}{U}\sum_{l=1}^{U/2} (Z_{jl} + Y'_{jl})
\end{equation}
\begin{equation}
    c_j^{(-i)} = \frac{1}{U-1}(\sum_{\substack{l=1 \\ l \neq i}}^{U/2}Z_{jl} + \sum_{l=1}^{U/2}Y'_{jl})
\end{equation}
where, $c_j^{(-i)}$ represents the centroid of the $j$-th speaker computed excluding the $i$-th utterance, and $Z_{jl}$ is the $l$-th student network embedding of the $j$-th speaker. 
The similarity matrix $S$ is defined as scaled cosine similarity between the student network embeddings and all centroids. 
\begin{equation}
S_{ji,k} = 
  \begin{cases}
  w \cdot cos(Z_{ji}, c_j^{(-i)}) + b \quad $if$ \ \ k = j \\
  w \cdot cos(Z_{ji}, c_k) + b \qquad $otherwise$, \\
  \end{cases}
\end{equation}
where, $S_{ji,k}$ denote the scaled cosine similarity between the $j$-th speaker's $i$-th the student network embedding and $k$-th speaker's centroid; $w$ and $b$ are learnable parameters. 
The final GE2E-H loss is as follows: 
\begin{equation}
    L_{G_H} = \frac{exp(S_{ji,j})}{\sum_{ji}exp(S_{ji,k})}
\end{equation}
Along with the speaker identification training over the entire training dataset for additional discriminative power \cite{kye2020meta}, the loss function of the student network can be defined as: 
\begin{equation}
\label{eq}
    L_S = L_{G_H} + L_C, \\
\end{equation}
where, $L_C$ is the classification cost, specified as CCE. 
It also computes the symmetric $\tilde{L}_S$ by entering the opposite mini-batch to train the student network across the entire training set. 
$\tilde{L}_S$ is derived by feeding $m'$ to the student network and $m$ to the MT network as in \cite{grill2020bootstrap}. 
The final loss function of the proposed method is denoted as follows:
\begin{equation}
    L = (L_S + \tilde{L}_S) / 2.
\end{equation}

\section{Experiments \& Result}
\label{sec:exp}
  
\subsection{Dataset}
For experiments in this study, we used the VoxCeleb2 dataset \cite{chung2018voxceleb2} for training, and the VoxCeleb1 test set \cite{nagrani2017voxceleb} for evaluation. 
The VoxCeleb2 dataset comprises over 1 million utterances from 6,112 speakers, and the VoxCeleb1 test set comprises 4,874 utterances from 40 speakers. 
Every utterance is recorded in mono, 16 kHz sampling rate, and 16-bit resolution. 
In addition, we augment the input data using room impulse response simulation and MUSAN corpus \cite{snyder2015musan}.

\subsection{Experimental configurations}
We used the raw waveforms as input, and the mini-batch was configured by setting the length of the input utterance to 59,049 samples($\approx$ 3.69 s) in the training phase. 
In the testing phase, we applied a test time augmentation by sampling ten temporal crops at regular intervals, to make the length of an input utterance the same as in the training \cite{chung2018voxceleb2}. 

In baseline experiments, the AMSGrad optimizer \cite{reddi2019convergence} was used with a learning rate (LR) of $0.001$ decaying exponentially with every iteration. 
On the other hand, we used a LARS \cite{you2017large} optimizer with an LR of $3$ and cosine annealing LR policy \cite{loshchilov2016sgdr} with warm-up\cite{goyal2017accurate} for the first three epochs in the proposed system.\footnote{These are the optimal combination of learning hyper-parameters found through internal experimentation of each system.} 
We applied a  weight decay with $\lambda=1e^{-4}$. 

\begin{table}[t]
    \caption{Comparison experiments of the baseline system using various loss functions and batch size.}
    \label{tab:baseline_contra}
    \centering
    \begin{tabular}{l|c|c|c}
        \Xhline{1pt} 
        System & Loss & Batch size& EER \\
        \Xhline{1pt} 
        \textbf{\#1-Baseline} & \multirow{2}{*}{CCE} & 200 & \textbf{2.24}\\
        \#2 &  & 1,920 & 2.51\\
        \hline
        \#3 & \multirow{2}{*}{GE2E} & 200 & 3.19\\
        \#4 &  & 1,920 & 3.34\\
        \hline
        \#5 & \multirow{2}{*}{AP} & 200 & 2.9\\
        \#6 & & 1,920 & 3.02\\
        \hline
        \#7 & \multirow{2}{*}{GE2E + CCE} & 200 & 2.73\\
        \#8 &  & 1,920 & 2.7\\
        \hline
        \#9 & \multirow{2}{*}{AP + CCE} & 200 & 2.52\\
        \#10 & & 1,920 & 2.9\\
        \Xhline{1pt} 
    \end{tabular}
    \vskip -5pt
\end{table}

\subsection{Results}

Table \ref{tab:baseline_contra} depicts experiment results of the baseline using various loss functions and batch sizes. 
When comparing System \#1 and 3, 5, 7, 9, we observed that the models trained only with metric learning or combined with the CCE performed worse than the baseline trained only with the CCE loss function. 
In addition, although large batches are known to be effective in metric learning, when comparing Systems \#3, 5, 7, 9 and 4, 6, 8, 10, performance degradation occurred in the rest of the loss function combinations, except for the GE2E and CCE combinations. 
From these results, we interpret that a metric learning-based system is more difficult to train than a classification-based system, as argued in \cite{bai2020speaker}.  

Table \ref{tab:proposed} shows the performance of the proposed method under the various conditions. 
In the first row, NP refers to the use of negative pairs, BC indicates batch configurations, S represents the use of the same mini-batch with added noise as the conventional MT, and D refers to the mini-batch comprising different utterances of the each speaker. 
In addition, in the LT column, which represents the learning target of the student network, P and E refer to the prediction and embedding of the MT network, respectively. 
System \#1, which reflects the original MT framework training scheme, shows a noticeable performance decline compared to the baseline (2.24\% versus 4.98\%). 
Thus, it can be considered that the original MT framework is unsuitable for the supervised learning SV task. 
When comparing System \#1 to 4, learning speaker embeddings of MT with negative pairs is more effective than predictions, as is configuring a mini-batch with different utterances rather than using the same mini-batch with noise. 
However, the performance has not improved. 
The comparison of Systems \#4 to 7 demonstrated that cosine similarity-based metric learning can improve the generalization performance of SV. 
Finally, System \#7, our proposed method, reported an EER of 1.98\%, with an RER of 11.61\% compared with the baseline. 

Table \ref{tab:batch} shows comparison results according to the batch size of the proposed method and the number of utterances per speaker. 
The batch size is the product of the number of speakers and the number of utterances per speaker included in a single mini-batch. 
Any batch size that is not an integer multiple of the utterance is set to the nearest integer number of speakers. 
Experimental results show that in the proposed framework, unlike the baseline system, large batch size is effective and reports the best performance when the batch size and number of utterances per speaker are 1920 and 4. 
These results can be posited as follows: the proposed method works stably when performing metric learning in large-scale batches using more accurate embedding from the temporal averaging model, MT. 

\begin{table}[t]
    \caption{Results of the proposed method for various training conditions. 
    (NP: Negative Pair, BC: Batch Configuration, S: Same batch, D: Different batch, LT: Learning Target, P: Prediction, E: Embedding)
    }
    \label{tab:proposed}
    \centering
    \begin{tabular}{l|c|c|c|c|c}
        \Xhline{1pt} 
        System & Consistency loss & NP & BC & LT &EER \\
        \Xhline{1pt} 
        \#1-Org\_MT & \multirow{4}{*}{MSE} & $\times$ & S & P &4.98\\
        \#2 &   & $\times$ & S & E &3.56\\
        \#3 & & $\times$ & D & E &2.37\\
        \#4 &  & \checkmark & D & E &2.28\\
        \hline
        \#5 & GE2E & \checkmark & D & E &2.27\\
        \hline
        \#6 & AP & \checkmark & D & E &2.18\\
        \hline
        \textbf{\#7-Proposed} & GE2E-H &\checkmark& D & E& \textbf{1.98} \\
        \Xhline{1pt} 
    \end{tabular}
    \vskip -3pt
\end{table}

\begin{table}[!t]
    \caption{Performance comparison of the proposed framework according to the batch size and the number of utterances per speaker. }
    \label{tab:batch}
    \centering
    \begin{tabular}{c|ccc}
        \Xhline{1pt} 
         & \multicolumn{3}{c}{\# Utterances per speaker} \\
         \cmidrule(lr){2-4}
        Batch size & 4 & 6 & 8\\
        \Xhline{1pt} 
        200 & 2.1 & 2.23 & 3.04\\
        800 & 2.12 & 2.07 & 2.19 \\
        1920&\textbf{1.98} & 2.09 & 2.28\\
        \Xhline{1pt} 
    \end{tabular}
    \vskip -5pt
\end{table}

\section{Conclusion}
\label{sec:conclusion}
We propose to apply the MT framework to improve the generalization performance of open-set SV tasks. 
The conventional MT framework is used in semi-supervised manner, with the same data for mini-batches but different augmentation for the teacher and student networks. 
In our study, we use MT network in supervised manner with mini-batches of different utterances from each speaker, increasing the cosine similarity of the embeddings from the two networks, which can decrease the intra-class variances. 
We also investigate a method to find discriminative embedding spaces suitable for open-set SV by utilizing diverse cosine similarity-based metric learning cost function with positive and negative embedding pairs. 
The proposed system showed an EER of 1.98\% on the VoxCeleb1 test set with an RER of 11.61\% compared to the baseline system, the improved RawNet2. 
In future work, we intend to conduct further experiments to demonstrate the superiority of the proposed method for SV tasks using encoders with different input features other than the raw waveforms. 

\section{Acknowledgements}

This research was supported and funded by the Korean National Police Agency. [Project Name: Real-time speaker recognition via voiceprint analysis / Project Number: PA-J000001-2017-101]

\bibliographystyle{IEEEtran}

\bibliography{mybib}

\end{document}